\documentclass[12pt,a4paper]{article}
\usepackage[top=3cm, bottom=3cm, left=3cm, right=3cm]{geometry}
\usepackage[latin1]{inputenc}
\usepackage{amsmath}
\usepackage{amsthm}
\usepackage{amsfonts}
\usepackage{amssymb}
\usepackage{graphics}
\usepackage[hang,flushmargin]{footmisc} 

\usepackage{amsmath,amsthm,amssymb,graphicx, multicol, array}
\usepackage{enumerate}
\usepackage{enumitem}
\usepackage{xcolor}
\usepackage{currfile}

\usepackage[bookmarks,colorlinks,breaklinks]{hyperref}

\hypersetup{linkcolor=blue,citecolor=red,filecolor=blue,urlcolor=blue} 

\usepackage{tabto}

\usepackage{tikz}
\usepackage{pgfplots}

\newtheorem{thm}{Theorem}[section]

\newcommand{\Z}{\mathbb{Z}}

\usepackage{tocloft}

\usepackage{fancyhdr}
\title{Deterministic Integer Factorization Algorithms}
\date{}
\author{N. A. Carella}

\begin{document}
\thispagestyle{empty}
\date{}

\maketitle
\textbf{\textit{Abstract}:} A new integer deterministic factorization algorithm, rated at arithmetic operations to $O(N^{1/6+\varepsilon})$ arithmetic operations, is presented in this note. Equivalently, given the least $(\log N)/6$ bits of a factor of the balanced integer $N = pq$, where $p$ and $q$ are primes, the algorithm factors the integer in polynomial time $O(\log(N)^c)$, with $c \geq 0$ constant, and $\varepsilon > 0$ an arbitrarily small number. It improves the current deterministic factorization algorithm, rated at arithmetic operations to $O(N^{1/5+\varepsilon})$ arithmetic operations. \let\thefootnote\relax\footnote{ \today \date{} 
}
\tableofcontents
\section{Introduction} \label{SF8888}
Let $\mathcal{A}=\{N=pq:p^{1+\delta}<q<ap^{1+\delta}\}$ and  $\mathcal{B}=\{N=pq:p<q<ap\}$, where $p$ and $q$ are primes, and $a > 1$, and $\delta > 0$ are constants, be the subsets of unbalanced integers, and balanced integers respectively. The factorization of a balanced integer has the worst time complexity, while the factorization of an unbalanced integer does not have the worst time complexity. Thus, the time complexity of 
integer factorization is measured by the time complexity of factoring balanced integers. This article discusses a new deterministic integer factorization algorithm acting on the subset of balanced integers in details. The running time complexity of this algorithm is deterministic exponential time $O(N^{1/6+\varepsilon})$, where $\varepsilon > 0$ is a small number. This improves the current performances of deterministic integer factorization algorithms rated at the running time complexities $O(N^{1/4+\varepsilon})$. The standard references \cite{CO1993}, \cite{CP2006}, \cite{LA2000}, \cite{MV1997}, \cite{RH1994}, \cite{SV2005}, \cite{WH1998} 
and others, provide extensive details on the theory of integer factorizations.\\

The innovation here involves a technique for generating multivariables polynomial equation $f(x,y)=(a_1x+a_0)(b_1y+b_0)\in \Z[x,y]$ for the integer factorization problem, the earlier work in \cite{CD1997} use a simpler polynomial $f_0(x,y)=(x+a_0)(y+b_0)\in \Z[x,y]$. The main result is the following.

\begin{thm}\label{thmF8888.100} Any integer $N$ can be decomposed into its prime factors in deterministic time complexity $O(N^{1/6+\varepsilon})$, where $\varepsilon > 0$, is an arbitrary small constant.
\end{thm}

Equivalently, given the least $(\log N)/6$ bits of a factor of $N = pq$, $p$ and $q$ primes, the algorithm factors the integer in polynomial time $O(\log(N)^c)$, with $c \geq 0$ constant. This is an improvement of the Coppersmith algorithm, which requires the least $(\log N)/4$ bits, see Theorem \ref{thmF2222.200}.\\

The proof of this result follows from Theorem \ref{thmDF4444.600} in Section \ref{SF4444}. Section \ref{SF2222} has a survey of the current deterministic integer factorization algorithms, and a new analysis of the Fermat Method. Section \ref{SF3333} covers some information on polynomials equations.

\section{Deterministic Exponential Time $O(N^{1/4})$}\label{SF2222}
There are many integer factorization algorithms of deterministic exponential time complexities $O(N^{1/4+\varepsilon})$, where $ \varepsilon > 0$ is an arbitrary small number, in the literature, see \cite[p. 238]{CP2006}, \cite{CO1993}, \cite{MP1996}, et cetera.\\

Integers $N=pq$ composed of nearly equal factors $p$ and $q=p+d$, where $d$ is small, can be factored by completing the square $N+d^2/4=x^2$. 
This produces the factor $p=\sqrt{N+d^2/4}-d/2$. For example, a factor of a product of any twin primes pairs is $p=\sqrt{N+1}-1$. This is one of the
earliest and most important integer factorization algorithms. It is usually called the Fermat factoring method, and uses an equivalent formulation 
based on the difference of squares equation $4N = x^2- y^2$. The Fermat factoring method effectively handles any integer $N \geq 1$ with a pair of 
factors in the ranges
\begin{equation} \label{el21}
N^{1/2}-N^{1/4} \log^c N<p <N^{1/2}  \qquad \text{ and } \qquad N^{1/2}<q <N^{1/2}+N^{1/4} \log^c N,
\end{equation}
with $c \geq 0$ constant. A brute force search for the prime factors $p$ and $q$ of the integer $N = pq$ in the ranges (\ref{el21}) has exponential time complexity of approximately $O(N^{1/4}\log^{c} N)$ arithmetic operations, since
\begin{equation} \label{el22}
	q=\frac{N}{[N^{1/2}]+x} \qquad \text{ for some integer} \qquad 0 \leq x\leq N^{1/4} \log^c N.
\end{equation} 
The Fermat factoring method miraculously reduces the running time complexity from exponential time complexity 
$O(N^{1/4}\log(N)^c )$ to polynomial time complexity $O(\log^c N)$. This is accomplished by computing a solution $U=p+q$, and $V=q-p$ of the equation $4N=U^2-V^2$, instead of directly and independently computing the prime factors $p$ and $q$. The basic structure of this method is attributed to Fermat, but it is not clear if the time complexity analysis was known before modern time. Statement (i) is well known, \cite[p. 148]{RH1994}, \cite[p. 19]{DB2002}, and other sources. But statement (ii) 
seems to be new. Here the symbol $[ x ]$ denotes the largest integer function. 
\begin{thm} \label{thmF2222.100} Suppose that an integer $N=pq$ has a pairs of prime factors $p$ and $q$. Then, the followings hold.
\begin{enumerate}[font=\normalfont, label=(\roman*)]
\item If the factors $p$ and $q$ satisfy $$|N^{1/2}-p|=O(N^{1/4}\log^c N)\qquad \text{ and }\qquad |N^{1/2}-q|=O(N^{1/4}\log^c N),$$ 
then it can be factored in deterministic 
polynomial time complexity $O(\log(N)^c)$, with $c \geq 0$ constant.
\item If the factors $p$ and $q$ satisfy $$|N^{1/2}-p|=O(N^{1/2} \log^c N)\qquad \text{ and }\qquad|N^{1/2}-q|=O(N^{1/2} \log^c N),$$ 
then it can be factored in deterministic 
exponential time complexity \\
$O(N^{1/4} \log(N)^{2c})$, with $c \geq 0$ constant. 
\end{enumerate}
\end{thm}

\begin{proof}[\textbf{Proof}] \textbf{Case (i).} To prove this claim, consider the approximations
\begin{equation} \label{eq2323.100}
p=[N^{1/2}]+x \qquad  \text{ and } \qquad  q=[N^{1/2}]+y,
\end{equation}
where $0 \leq | x |, | y |= O(\log(N)^c)$, and $c \geq 0$ is a constant. The search  for a solution $z=x+y$ of the equation $4N=V^2-V^2$, where $U=p+q$ and $V=q-p$, start with the sequence of approximations
\begin{equation} \label{eq2323.110}
	U_z=[N^{1/2}]+x+[N^{1/2}]+y=2[N^{1/2}]+z ,
\end{equation}
where $z= 0, \pm 1, \pm 2, \ldots $. The approximate number of cycles required to determine a solution $(U_z, V_z)$ is at most\\
\begin{eqnarray} \label{eq2323.120}
p+q- \left (2N^{1/2}+z+\frac{N}{2N^{1/2}+z} \right ) 
	&\leq& \frac{\left (N^{1/2}-p \right )^2}{p}  \\ &=&O(\log ^{2c}N)\nonumber
\end{eqnarray}
cycles. Therefore, the time complexity of the algorithm is at most $O(\log^{2c} N)$ arithmetic operations.\\

\textbf{Case (ii).} To prove this claim, consider the approximations
\begin{equation} \label{eq2323.200}
	p=[N^{1/2}]+[N^{1/4}]x_0+x \qquad  \text{ and } \qquad  q=[N^{1/2}]+[N^{1/4}]y_0+y,
\end{equation}
where 
\begin{enumerate}
\item $\displaystyle 0 \leq | x_0 |, | y_0 |= O(N^{1/4}),$
\item $\displaystyle 0 \leq | x |, | y |= O(\log(N)^c),$
\end{enumerate}
and $c \geq 0$ is a constant. The search  for a solution $z=x+y$ of the equation $4N=V^2-V^2$, where $U=p+q$ and $V=q-p$, start with the sequence of approximations
\begin{eqnarray} \label{eq2323.210}
	U_z&=&[N^{1/2}]+[N^{1/4}]x_0+x+[N^{1/2}]+[N^{1/4}]y_0+y\\
	&=&2[N^{1/2}]+2[N^{1/4}]z_0+z\nonumber ,
\end{eqnarray}
where $ 0 \leq | x_0 |= O(N^{1/4})$ is a given parameter, and $z= 0, \pm 1, \pm 2, \ldots $. Given the correct parameter $z$, the approximate number of cycles required to determine a solution $(U_z, V_z)$ is at most\\
\begin{eqnarray} \label{eq2323.220}
p+q- \left (2N^{1/2}+[N^{1/4}]z_0+z \right )
	&\leq& \frac{\left (N^{1/2}+[N^{1/4}]z_0-p \right )^2}{p}  \\ &=&O(\log ^{2c}N)\nonumber
\end{eqnarray}
cycles. Since the correct parameter $| z_0 |= O(N^{1/4})$, the time complexity of the algorithm is at most $O(N^{1/4}\log^{2c} N)$ arithmetic operations.
\end{proof}

Algorithms that compute multiples of 
\begin{equation} \label{eq2323.250}
p + q, \quad q- p,\quad ap + bq, \quad aq-bp,
\end{equation}
are the topic of current research in integer factorization and related topics, consult \cite{LR1974}, \cite{HH2021}, et alii. The Pollard factoring method, and the elliptic curve factoring method, and a few other algorithms are based on the direct or indirect calculations of 
multiples of $p+q=N+1-\varphi(N)$, or $q-p= \sqrt{(p+q)^2-4N}$. It should be noted that multiples of $p^2+q^2=N^2+1-\varphi_2(N)$, are also 
effective.\\

Another related, and recently discovered integer factorization algorithm in this class is the following.

\begin{thm} \label{thmF2222.200} {\normalfont (\cite{CD1997})} If the least (or most) significant $(\log N)/4$ bits of a prime factor $p$ or $q$ of the integer $N = pq, p < q < 2p$, 
are given, then it can be decomposed in deterministic polynomial time complexity $O(\log(N)^c)$, $c > 0$ constant.
\end{thm}

An improved version of this result is given in Theorem \ref{thmF8888.100}. The Fermat factoring algorithm (Theorem \ref{thmF2222.100}), and the Coppersmith factoring algorithm (Theorem \ref{thmF2222.200}), are equivalent integer factorization algorithms of the same running time complexity $O(N^{1/4}\log(N)^c)$. Moreover, both have 
equivalent proofs based on lattice reduction methods. Both of these algorithms are practical for small integers, for example, $N = O(2^{200})$ or 
thereabouts. Some improvement to the Fermat method is reported in \cite{EG2009}, and \cite{MJ1999}, and experimental data for the Coppersmith factoring algorithm 
are compiled in \cite{CJ2007}, \cite{EJ2005}, \cite{HJ2008}, and many other similar references.\\

Another class of algorithm, based on efficient evaluations of high degree polynomials, is stated below. The author
 of this paper also have a survey of current literature on this class of integer factorization algorithms.
 
\begin{thm} \label{thmDF2222.300} {\normalfont (\cite{CH2013})} There exists a deterministic algorithm that computes the prime factorization of a positive integer $N$ in 
$O(N^{1/4}\log(N) (\log \log(N)^{-1/2})$ bit operations.
\end{thm}

\section{Deterministic Exponential Time $O(N^{1/5})$}\label{SF2255}
The fastest, deterministic, and unconditionally proven integer factorization algorithms in the literature have running time complexities $O(N^{1/5+\varepsilon})$. This is a very recent development. The previous algorithm of the same complexity was conditional on the GRH. This conditional integer factorization algorithm is based on the class number of numbers fields, see \cite[p. 251]{CP2006} for some details. 

\begin{thm} \label{thmF2255.100} {\normalfont (\cite{HH2021})} There is a deterministic integer factorization algorithm achieving $O(N^{1/5+\varepsilon})$ arithmetic operations.
\end{thm}


\section{Basic Systems Of Polynomials Equations}\label{SF3333}
The applications of lattice reduction methods to the theory of polynomial equations and its applications to cryptography are considered in fine 
details in \cite{VT1989}, \cite{CD1997}, \cite{HN1997}, \cite{BM2005}, \cite{CJ2007}, \cite{CT2013},  \cite{JF2007}, \cite{LH2008} and others.\\

Comprehensive introductions to lattice reduction methods and its applications to polynomials equations are given in \cite[Chapter 2]{DG2002}, \cite[Chapter 3]{JE2007},
 \cite[Chapter 3]{MA2003}, and similar sources. The evolving analysis on a few specific polynomial equations of three variables is given in \cite{BJ2007}.\\
 
Employing lattice reduction methods, several results for the polynomials 
\begin{equation}
f(x,y)=\sum_{0\leq i,j \leq d}a_{i,j}x^iy^j
\end{equation}
and
\begin{equation}
f(x,y,z)=\sum_{0\leq i,j,k \leq d}a_{i,j,k}x^iy^jz^k
\end{equation} 
of two and three variables respectively, have been unconditionally proven. A relevant result 
from the theory of polynomials equations is included in this Section.

\begin{thm}\label{thmF3333.100}{\normalfont(\cite{CD1997})} Let $f(x, y) \in \mathbb{Z}[x, y]$ be an irreducible polynomial of maximum degree $\text{deg}(f)=d$ in each variable, and let $(x_0, y_0)$ be a root of $f(x, y) = 0$, such that $0 \leq | x_0 | \leq X$, $0 \leq | y_0 | \leq Y$. The height of the polynomial $f(xX, yY)$ 
is defined by
\begin{equation}
W=\left || f(xX,yY) \right ||_{\infty}=\max\{ |a_{i,j}X^iY^j|:0\leq i,j \leq d\}.
\end{equation}
\begin{enumerate}[font=\normalfont, label=(\roman*)]
\item If $XY < W^{2/(3d)}$, then the roots $(x_0, y_0)$ can be determined in deterministic polynomial time $O(\log ^cN$, $c > 0$ constant.
\item If $XY < W^{1/d}$, and the total degree of the polynomial satisfies $0 = i + j = d$, then the roots $(x_0, y_0)$ can be determined in 
deterministic polynomial time $O(\log^c N)$, $c > 0$ constant.
\end{enumerate}
\end{thm} 

The detailed heuristic analysis of the lattices for $f(x, y, z) = c_0xy + c_1x + c_2y + c_3z + c_4 \in \mathbb{Z}[x, y, z]$, and a few other 
polynomials of three variables and related applications, appear in \cite{BJ2007}, \cite[p. 8]{EJ2005}, and \cite[p. 66]{JE2007}. Practical applications also appear in \cite{JE2007} and
 \cite[p. 11]{HJ2008}.

\section{Deterministic Exponential Time $O(N^{1/6})$}\label{SF4444}
For a pair of fixed parameters $0 < \alpha < 1 < \beta$, let $p$ and $q$ be prime numbers such that $\sqrt{\alpha N}<p<\sqrt{N}$, and 
$\sqrt{N}<q<\sqrt{\beta N}$, let $\gamma \sqrt{N}=(\sqrt{\alpha N} +\sqrt{\beta N})/2$ be the arithmetic mean of the interval $[\sqrt{\alpha N}, \sqrt{\beta N}]$. The Fermat factoring 
method, (Theorem \ref{thmF2222.100}) and Coppersmith factoring method (Theorem \ref{thmF2222.200}) assume that the prime factors of the integer $N = pq$ are centered at the 
symmetric center $\sqrt{N}$ of the interval $[\sqrt{\alpha N}, \sqrt{\beta N}]$. Shifting the symmetric center $\sqrt{N}$ to the arithmetic mean 
center $\gamma \sqrt{N}$ of the factors, or to a pair of distinct centers $\sqrt{\alpha N}$ and $\sqrt{\beta N}$, with $\alpha,\beta\ne 1$, can be 
used to derive various multivariable polynomials, which reduce the time complexities of both the Fermat factoring method, and Coppersmith factoring 
method, respectively. 

\begin{thm}\label{thmDF4444.600} Given the least (or most) significant $(\log N)/6$ bits of a prime factor $p$ or $q$ of a large integer $N = pq, p < q < 2p$, the integer $N$ can be decomposed in deterministic polynomial time $O(\log ^c N)$, with $c > 0$ constant.
\end{thm}   

\begin{proof}[\textbf{Proof}] Suppose that $N = pq$ has balanced prime factors such that $\sqrt{N/2}<p<\sqrt{N}<q<\sqrt{2N}$ centered at the symmetric center 
$\sqrt{N}$ of the interval $[\sqrt{N/2},\sqrt{2N}]$, and the least significant $(\log N)/6$ bits $x_0$ of the prime factor $p$ are given. Here the symbol $[ x ]$ denotes the largest integer function. Here the symbol $[ x ]$ denotes the largest integer function.\\

Let $B=[N^{1/6}]$, and let $B^3=[N^{1/2}]$. Assume the integer (or prime number) $B>x_0$ satisfies the condition $\gcd(B, x_0)=1$. Now, expand the factors as $B$-adic integers
\begin{equation}\label{eqDF4444.610}
p=x_1B^3+B^2x_3+Bx_2+x_0=x_1B^3+B(Bx_3+x_2)+x_0,
 \end{equation}
and 
\begin{equation}\label{eqDF4444.620}
q=y_1B^3+B^2y_3+By_2+y_0=y_1B^3+B(By_3+y_2)+y_0,
\end{equation}
where the variables have the following dynamic ranges.
\begin{enumerate}  
\item $0 \leq |x_i|,|y_i| < N^{1/6}$, \tabto{10.8cm} for $i\in \{0,1,2,3\}$,
	
\item $0 \leq |Bx_3+x_2|=x \leq N^{1/3}$, 

\item $0 \leq |By_3+y_2|=y \leq N^{1/3}$,
	
\item $x_1y_1\leq \log N$. 
\end{enumerate}
Lines 2 and 3 show the changes of variables $x=Bx_3+x_2$ and $y=By_3+y_2$. The last condition is arises from $x_1y_1B^6=x_1y_1[ N^{1/6}]^6\leq N$, for balanced factors $p<q<2p$, the small variables $x_1, y_1\in \{1,2\}$ work well.\\

Proceed to use the given least significant $(\log N)/6$ bits $x_0<B$ of the prime factor $p$ to compute $y_0<B$ via the congruence equation
\begin{equation}\label{eqDF4444.630}
	\left ( x_1B^3+Bx+x_0\right)\left ( y_1B^3+By+y_0\right)-N \equiv 0 \text{ mod } B 
\end{equation}

By the initial conditions $\gcd(B,x_0)=1$ on the integer $B$, this congruence has a unique solution 
\begin{equation}\label{eqDF4444.640}
	y_0 \equiv N\cdot x_0^{-1} \bmod B.
\end{equation}
Next, expanding the product $N=\left ( x_1B^3+Bx+x_0\right)\left ( y_1B^3+By+y_0\right)$ yields the corresponding polynomial
\begin{eqnarray}\label{eqDF4444.650}
f(x,y)&=&\left ( x_1B^3+Bx+x_0\right)\left ( y_1B^3+By+y_0\right)-N\\
&=&B^2xy+\left(y_1B^3+y_0 \right )Bx+ \left(x_1B^3+x_0 \right )By\nonumber \\
&& \hskip 2 in + \left ( x_1B^3+x_0\right)\left ( y_1B^3+y_0\right)-N\nonumber \\
&=&c_3xy+c_2x+c_1y+c_0\nonumber.
\end{eqnarray}
The coefficients are:
\begin{enumerate}\addtocounter{enumi}{4}
\item $c_0=\left ( x_1B^3+x_0\right)\left ( y_1B^3+y_0\right)-N$,

\item $c_1=x_1B^3+x_0$,

\item $c_2=y_1B^3+y_0 $,

\item $c_3=B^2$.
\end{enumerate}
To demonstrate that $f(x,y)\in \Z[x,y]$ is an irreducible polynomial, consider the factorization
\begin{equation} \label{eqDF4444.660}
f(x,y)=c_3xy+c_2x+c_1y+c_0=(a_1x+a_0)(b_1y+b_0)
\end{equation}
into linear factors. Matching coefficients yields
\begin{enumerate}\addtocounter{enumi}{8}  
\item $a_0b_0=c_0=\left ( x_1B^3+x_0\right)\left ( y_1B^3+y_0\right)-N$,

\item $a_0b_1=c_1=x_1B^3+x_0$,

\item $a_1b_0=c_2=y_1B^3+y_0 $,

\item $a_1b_1=c_3=B^2$.
\end{enumerate}
But, the constant terms
\begin{eqnarray}\label{eqDF4444.670}
 a_0b_0&=&\frac{(a_0b_1)\cdot(a_1b_0)}{a_1b_1}\\&=&\frac{\left ( x_1B^3+x_0\right)\left ( y_1B^3+y_0\right)}{B^2}\nonumber\\&\ne& c_0\nonumber
\end{eqnarray}
do not agree. Therefore, this is an irreducible polynomial  over the integers. \\

By construction, the polynomial $f(x,y)\in \Z[x,y]$ has a small root $(x,y)=(x_4,y_4)$ such that $p=x_1B^3+Bx_4+x_0$ is a factor of $N$.\\

To estimate the upper bounds of the solutions $x$ and $y$, it is sufficient to estimate the height $W=||f(yYxX)||_{\infty}$ of the polynomial $f(xX, yY)$. Suppose that $ X=Y=N^{1/3}$, and $B$ is prime. Then, 
\begin{equation}\label{eqDF4444.680}
\gcd(c_0,c_1,c_2,c_3)=1,
\end{equation}
where $c_0\equiv c_3\equiv 0 \bmod B$, but $c_1, c_2\not \equiv 0 \bmod B$. Thus, it follows that the height is given by
\begin{equation}\label{eqDF4444.685}
||f(yYxX)||_{\infty}=\max\{c_3XY,c_2X,c_1Y,c_0\} = N,
\end{equation}
for example, $c_3XY=B^2XY=N^{1/3}XY=N$.\\

Since $XY \leq W^{2/3d}=N^{2/3}$, where $\text{deg}(f)=d=1$ is the total degree of the polynomial, using lattice reduction methods, the small integer roots $x_1 \leq N^{1/3}$ and $x_1 \leq N^{1/3}$ can be determined in deterministic polynomial time, see Theorem \ref{thmF3333.100}.
\end{proof} 

\textbf{Note 1.} Simple modification of Theorem \ref{thmDF4444.600} can be used to handle all the other factorizations of nonbalanced integers $N = pq$ with $p$ and $q$ primes such that 
$(\alpha N)^{1/2-\delta}<p<N^{1/2}<q< (\beta N)^{1/2+\delta}$ with $\delta>0$. For example, if $(\alpha N)^{1/3}<p<N^{1/3}$ and $N^{1/3}<q<(\beta N)^{2/3}$, 
where $0<\alpha<\beta$ are small constants. Let $B=N^{1/6}$, and write the prime factors in the form
\begin{equation}
p=x_1B^2+Bx+x_0  \qquad \text{ and } \qquad q=y_1B^4+By+y_0,
\end{equation}
where $0 \leq |x| \leq N^{1/12}$, $0 \leq |y| \leq N^{7/12}$, $0 \leq |x_0|,|y_0| \leq N^{1/6}$, and $x_1y_1[N^{1/12}]^{12}\leq N$. Now proceeds as before, but making the 
necessary changes as needed. At most a few changes of parameter $\delta \in \{0,1/6,1/4, \ldots \}$ are required to cover all possible prime 
factorizations.\\


\currfilename.\\

\end{document}